# A 7-core erbium doped double-clad fiber amplifier with side-coupled pumping

Kazi S. Abedin,* John M. Fini, Taunay F. Thierry, Benyuan Zhu, Man F. Yan, Lalit Bansal, Frank V. Dimarcello, Eric M. Monberg and David J. DiGiovanni

*OFS Laboratories, 19 Schoolhouse Road, Somerset, New Jersey 08873*
*Corresponding author: kabedin@ofsoptics.com*

We demonstrate a 7-core erbium doped fiber amplifier employing side pumping using tapered multimode fiber. The amplifier has multicore inputs and outputs which can be readily spliced to multicore transmission fiber for amplifying space division multiplexed signals. Gain over 25dB was obtained in each of the cores over a 40-nm bandwidth covering C-band.
http://dx.doi.org/10.1364/OL.99.099999

Space division multiplexing (SDM) has drawn significant amount of interests lately as a potential means for increasing the capacity of optical transmission systems. SDM can be achieved by using fibers with multiple numbers of cores, and/or by using cores supporting multiple number of modes. Optical transmission at the bit rates over 100 Tb/s to 1 Pb/s has been demonstrated by using multicore and multimode fibers[1-5]. To overcome the loss in SDM transmission systems, suitable amplifiers that can amplify signals carried by all the SDM channels simultaneously will be needed.[6,7]

To amplify signals from multicore SDM systems, different forms of erbium doped fiber amplifiers (EDFAs) using multicore[8-11], bundled[12] and multi-element gain fibers[13] have been developed. Such amplifiers have already found applications in compensating losses in SDM transmission links[14-15].

A multicore EDFA can be pumped either through the core or the cladding. In core-pumped multicore EDFA[8,9], all the cores are pumped separately using single mode pump diodes, whereas in cladding pumped EDFAs, all of these cores are pumped simultaneously through a common multimoded cladding[10,11]. The latter scheme requires far fewer number of optical components and allows for using low-cost high power multimode pump diodes, which can potentially reduce manufacturing cost. Moreover, high power multimode pump source can be operated without using a Peltier temperature controller, with the promise of reduced energy consumption.

We have recently demonstrated operation of a multicore amplifier employing cladding pumping through one end of the gain fiber. Signals from single mode fibers are coupled into and out of the multicore gain fiber through fan-out devices (tapered fiber bundled couplers). Peak gain over 32 dB has been obtained at a wavelength of 1560 nm and the bandwidth measured at 20-dB gain was about 35 nm[10]. It is, however, highly desired for the MC amplifiers to have multicore input and outputs, so that MC amplifiers can be conveniently spliced to MC-transmission fibers in order to further reduce the cost and to simplify the amplifier design.

Besides end-face pump coupling, it is also possible to couple pump radiation through the side of the double-clad gain fiber. One remarkable advantage of side-coupling is that pump radiation can be coupled without interruption of the core waveguide. Side-coupling with high efficiency has been demonstrated in single core double-clad fibers in various bulk forms, such as V-groove side-coupling combiners[16], embedded-mirror combiners[17], direct fusion of tapered multimode fiber[18,19], and also in the form of distributed coupling using GT-wave fiber assembly.[20,21].

In this paper, we demonstrate a cladding-pumped multicore EDFA in which pump radiation is coupled to the gain fiber through the side of the cladding. The amplifier has a multicore input and output fiber, which can be readily spliced to multicore passive fibers. We have studied the amplification and noise properties of the amplifier in the C and L band region. An internal gain greater than 25 dB has been measured over 40 nm for all seven cores.

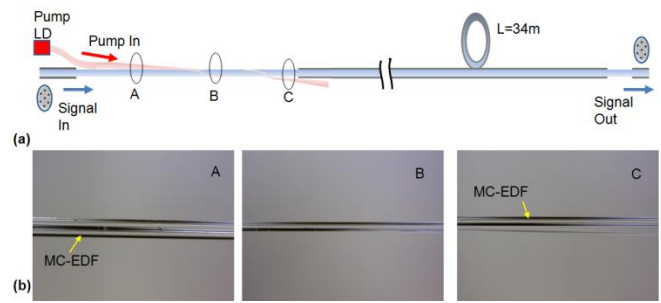

**Fig. 1.** A 7-core EDFA with side pumping. (a) Schematic diagram, (b) Photographs of the pump/signal combiner, taken near down-taper region (A), uniform diameter region (B), and up-taper region (C).

The schematic of the 7-core fiber amplifier with side coupled pumping is shown in Fig 1. The multicore erbium doped fiber had 7 identical cores that were arranged in a hexagonal array with a pitch of about 41 μm. The diameter of the cores and the numerical aperture were 3.2 μm (MFD: 5.6 μm) and 0.23, respectively. ,The fiber had a cladding diameter of 100 μm, and was coated with low index-coating for cladding pumping with a numerical aperture of 0.45. The absorption of the erbium-doped core for core-guided light at 1530 nm was ~6 dB/m. We measured the background losses at 1305 nm, for the central and outer cores using an OTDR and found them to be about the same, ~41dB/km.

In multicore fiber, since the outer cores are located close to the edge of the cladding, V-groove or embedded mirror methods are difficult to implement. To couple pump radiation into our fiber, we used a side-coupling method based on tapered multimode fiber. Side-coupling based on taper multimode fiber[18, 19] ensures higher average pump intensity along the length of the gain fiber. A short section of the gain fiber (~8 cm long) near the input end was stripped of its low-index coating, and the exposed section was made to have optical-contact with a tapered multimode pump fiber. The tapered fiber was fabricated by a process similar to that reported in Ref. 19, which involves slowly varying the diameter of the multimode pump fiber (105/125, NA=0.15) from 125 to 15 μm over 25 mm, and maintaining a uniform diameter section of length 20 mm followed by an up-tapered section. The tapered and uniform diameter sections were wound around the gain fiber by one and a half-turns to ensure efficient tacking, while the uptapered section was left detached from the gain fiber. Fig 1b shows photographs taken at different sections of the coupler for side pumping. Upon entering the down-tapered section, the pump radiation experiences an increase in NA[22,23], thus escaping the pump fiber and is gradually captured by the gain fiber which has higher NA. By launching 7 W pump power at 980 nm using 105/125 fiber, we could couple 4.7 W into the 100-μm gain fiber, corresponding to an efficiency of ~67%. Further enhancement in coupling is expected through optimization of the taper ratio, the taper length and by using coreless intermediary fiber[19]. The length of the section of the gain fiber between the input end and the pump-coupler that remained unpumped was ~25 cm.

When amplifying in a cladding pumped amplifier employing long length, the signal intensity can become large when compared with the pump intensity. This can deplete the upper state population, which makes it harder to achieve gain in the C-band with low noise figure, especially below 1560 nm, as we had observed in a cladding pumped multicore EDFA made of 50 m long gain fiber. Results of our numerical simulation[10] showed that the gain spectrum can be extended to C-band by choosing a shorter length erbium-doped fiber. In the current amplifier design we chose 34-m-long gain fiber.

The amplifcation and noise properties in each core of the 7-core amplifier were studied one at a time by coupling signal into and out of the gain fiber using two antireflection-coated lensed fibers attached to precision stages. To suppress the Fresnel reflection from the end facet, the gain fiber were angle-cleaved on both ends. In addition, optical isolators were used to suppress spurious backreflection at the input and output sides (connected to the pigtails of the lensed fibers).

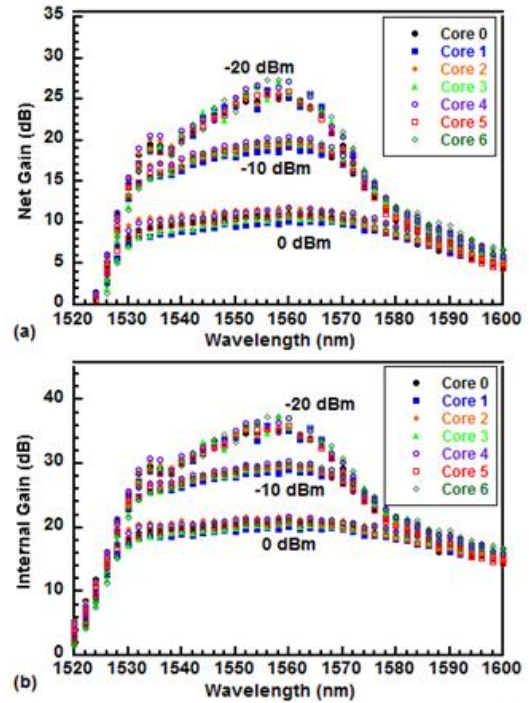

**Fig. 2.** Net (external) gain (a), and gross (internal) gain (b) measured for the 7 different cores of the MC-EDFA, for input signal powers of 0, -10, and -20dBm. The launched pump power was 4.7 W.

Figure 2(a) shows the net gain versus wavelength measured in the seven cores for input signal powers of -20, -10 and 0 dBm, when the pump power (coupled) was 4.7 W. Here the central core is represented as core 0. Small signal gain over 25 dB could be obtained near 1560 nm. Due to the angle cleaving, the coupling loss between the lensed fiber and gain fiber was rather high, estimated to be about 4.3 dB. The internal or gross gain is higher than the net gain by an amount equal to the total passive loss of the lensed fiber coupler and isolator. The gross or internal gain is plotted as a function of wavelength for the different cores in Fig 2(b). This gross gain is the expected gain that signals from a multicore transmission fiber will experience, when spliced directly to the multicore gain fiber. The maximum gross gain was about 32 dB near 1560 nm, and gain over 25 dB was obtained over a bandwidth of ~40 nm. We observed a significant expansion of the bandwidth into the C-band for this amplifier with shorter gain fiber than that used in our previous demonstration[10].

The internal NF was calculated from the gross gain and the ASE noise estimated at the output of each cores following the treatment as described in Ref. 8. The internal NF for different cores as a function of wavlength is shown in Fig. 3(a) and (b) for input signal powers of 0 and -10 dBm. As shown in Fig. 3(a) and 3(b), the NF becomes high for wavelength shorter than 1540 nm, while it tends to decrease for longer wavelengths. The NF was ~5 dB for the signal wavelength of 1560 nm. At 1530 nm, the NF was ~8 dB, much lower than the 15 dB NF observed previously for L = 50 m.

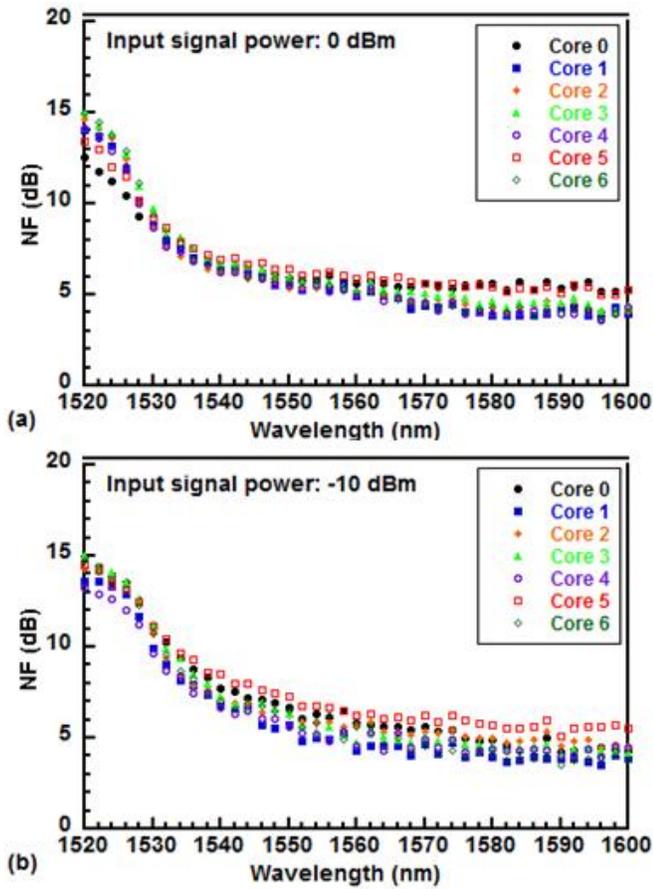

**Fig. 3.** Internal NF plotted as a function of wavelength for different cores of the multicore EDF amplifier. The coupled pump power was 4.7 W and the input signal power was 0 dBm in (a), and -10 dBm in (b).

We have performed numerical simulation, as described in Ref. 10, to estimate the gain and NF of a 7-core cladding pump EDFA with a length of 34 m and a pump power of 4.7 W. As shown in Fig. 4(a), an ~35 dB internal gain is expected for small-signals (-20 dBm) near 1560 nm, which is in good agreement with the experimental results. Figure 4(a) also shows the gain calculated for a 50 m-long gain fiber. For shorter wavelengths (<1540 nm), higher gain is obtained for the amplifier with L = 34 m.

The NF was also calculated for gain fiber with lengths of 34 m and 50 m for different signal wavelengths. As shown in Fig. 4(b), the NF is reduced significantly for shorter wavelengths as the length of the gain fiber is reduced. For longer signal wavelength >1560 nm, the NF remained relatively unchanged with fiber length. Since a significant amount of pump remains unabsorbed, there is negligible pump-depletion induced cross talk among the cores.

Since all the cores are pumped by a common source in cladding-pumped multicore amplifiers, controlling the gain of individual cores cannot be achieved by controlling the pump diode current. Individual core gain control might be possible by using suitable spatial variable optical attenuators (VOAs) at the amplifier output to control the signal intensity. Mulitcore amplifier design also require multicore fiber isolators for suppression of ASE and spurious backreflections. Realization of such multicore components is a topic for research in the near future.

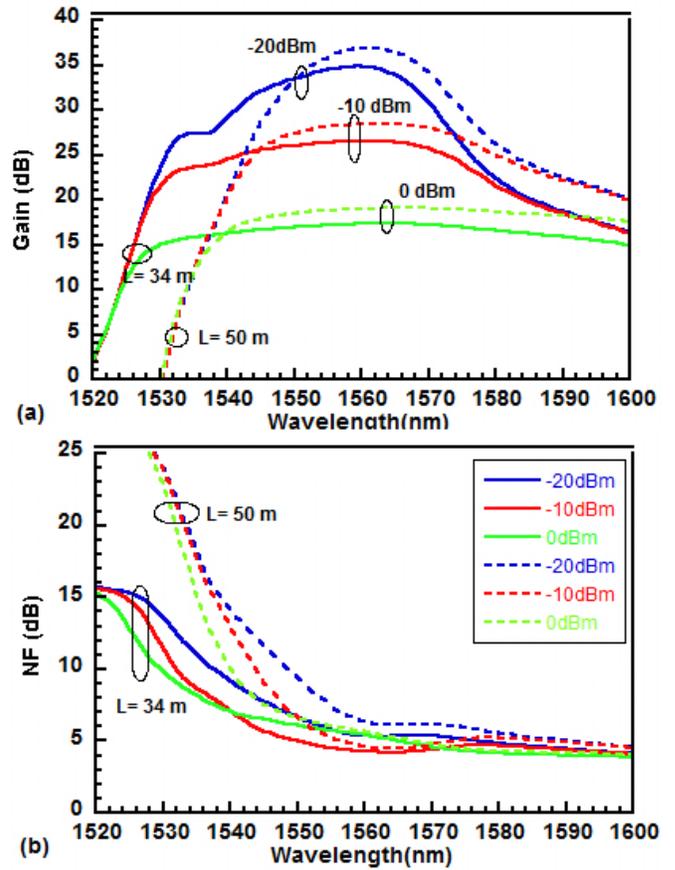

**Fig. 4.** Calculated gain (a) and noise figure (b) plotted as a function of wavelength for different input signals with power of 0 dBm, -10 dBm and -20 dBm. Pump power was 4.7 W.

In conclusion, we have demonstrated a 7-core cladding-pumped erbium-doped multicore fiber amplifier in which the pump was coupled through the side of the gain fiber. A peak net gain of about 32 dB was obtained at 1560 nm, with a NF of ~5 dB. Gain over 25 dB was observed over a bandwidth of about 40 nm centered around 1560 nm. The amplifier has multicore inputs and outputs which could be readily spliced to passive MCF with similar core-to-core pitch.